\documentstyle[eqsecnum,aps]{revtex}
\begin{document}
\draft
\title{A NON-ABELIAN VARIATION ON THE SAVVIDY VACUUM\\
         OF THE YANG-MILLS GAUGE THEORY\footnotemark[1]}
\footnotetext[1]{This work is supported in part by funds
provided by the U. S. Department of Energy (D.O.E.) under
contracts \#DE-FG06-88ER40427, \#DE-AC02-76ER03069
and \#DE-FG02-91ER40676}
\author{Suzhou~Huang}
\address{Department of Physics, FM--15,
         University of Washington\\
         Seattle, Washington ~98195\\
         {\rm and}
         Center for Theoretical Physics,
         Laboratory for Nuclear Science,
         and Department of Physics\\
         Massachusetts Institute of Technology,
         Cambridge, Massachusetts 02139}
\author{A.~R.~Levi}
\address{Department of Physics, Boston University\\
         590 Commonwealth Avenue, Boston, MA 02215}
\maketitle
\begin{abstract}
\ \\
\ \\
BU-HEP-93-22 \\
CTP 2267 \\
(December 1993) \\
\ \\
  As a prelude to a truly non-perturbative evaluation
of the effective potential in terms of lattice QCD,
the one loop effective potential for a non-Abelian gauge
configuration is calculated using the background field method.
Through a non-trivial correlation between the space and color
orientations the new background field avoids the possible
coordinate singularity, ${\rm Det}B_i^a=0$, observed recently
by Ken Johnson and his collaborators in their Schr\"{o}dinger
functional study of the SU(2) Yang-Mills theory. In addition,
since our ansatz generates a constant color magnetic field
through the commutator terms rather than derivative terms,
many of the technical drawbacks the Savvidy ansatz suffers
on a lattice can be avoided. Our one loop study yields
qualitatively the same result as that of Savvidy's.
\end{abstract}

\vfill
\eject
\widetext
\section{INTRODUCTION}

  Even after many years of intensive efforts, the vacuum
structure of non-Abelian gauge theories still remains
elusive to us. Due to the lack of other systematic
methods, the conventional approach is to analyze the
effective action using the so-called background
field method. In their pioneer work \cite{savvidy}
Savvidy and Matinyan, analogous to the case of QED,
calculated up to one loop order the effective potential
for an Abelian background gauge field,
\begin{equation}
       A^a_\mu(x) = {1\over2} H \delta^{a3}
       (x_1 \delta_{\mu 2}  - x_2 \delta_{\mu 1}  )\, ,
\label{abel}
\end{equation}
which generates a constant color magnetic field
$B_i^a=H\delta_{i3}\delta_{a3}$
in the SU(2) Yang-Mills theory. It was then found,
remarkably, that the vacuum with this background field
is energetically favored over the perturbative vacuum
($H=0$). Unfortunately, a more careful analysis
\cite{nielsen} soon revealed that there exists an
imaginary part in the effective potential, indicating
that the background field Eq.(\ref{abel}) is not a
minimum but rather a saddle point in the configuration
space in the context of the loop expansion.
The subsequent work by the Copenhagen group
\cite{copenhagen}, still within the framework of
loop expansion, tried to remedy the physically
appealing picture of Savvidy's vacuum by introducing
inhomogeneouty, leading to the so-called Copenhagen vacuum.

  However, the validity of the loop expansion has
been questioned by Maiani et al \cite{maiani}.
These authors strongly argued that the
calculation of the effective potential in the
presence of a background field is actually
non-perturbative. So far the only known non-perturbative
technique in field theory that can be made systematic
is the lattice approach.
More recently several works with the lattice approach
have appeared in the literature, both in four dimensions
\cite{ambjorn} and three dimensions \cite{woloshyn}.
Due to various technical reasons, mainly related to
the fact that the Savvidy ansatz in Eq.(\ref{abel})
is non-uniform, all these lattice works have yet
to yield a conclusive result.

  Recently, in an unrelated but very interesting
work by Johnson and his collaborators \cite{johnson},
the Schr\"{o}dinger functional approach in terms of
magnetic field strength is applied to the SU(2)
Yang-Mills gauge theory.
In the explicit Schr\"{o}dinger functional
Hamiltonian they find that there is a factor
$1/{\rm Det}B_i^a$ in the kinetic energy term, similar
to the $1/r$ factor in the quantum mechanics
Schr\"{o}dinger equation in polar coordinate.
This factor may signal a potential coordinate
singularity for the color magnetic field in the
configuration space, similar to that of
the wavefunctions in quantum mechanics have to
satisfy certain boundary conditions at $r=0$.
Since the Savvidy ansatz yields
${\rm Det}B_i^a=0$, due to its intrinsic Abelian nature,
it is desirable to seek an alternative background field,
which avoids this potential coordinate singularity.
A simple choice for SU(2) gauge group, suggested to us
by Ken Johnson, is the following
\begin{equation}
      A_0^a(x)=0,\,\,\,\,\, {\rm and}
      \,\,\,\,\, A_i^a(x)=h\delta_i^a\, ,
\label{nonabel}
\end{equation}
where $h$ is a constant in space-time. The corresponding
color magnetic field is simply given by
$B_i^a=gh^2\delta_i^a$ and
obviously has ${\rm Det}B_i^a\ne 0$ when $h\ne 0$.

  It is easy to recognize that the constant magnetic
field is generated by Eq.(\ref{nonabel}) through the
commutator terms in the field strength,
rather than the derivative terms as in the
Savvidy's case. It has been proved \cite{brown} that
there are no other ways to generate a constant
magnetic field. Due to the non-trivial correlation
between its directions in space and color space,
the background field Eq.(\ref{nonabel})
evades the possible coordinate singularity.
In addition, since the non-Abelian background is
space-time independent, it is
much easier to put it on a lattice than the non-constant
background field Eq.(\ref{abel}). For example, the
periodic boundary condition is automatic.

  Before we launch an extensive numerical simulation,
it is still worthwhile to calculate the one-loop
effective potential for the non-Abelian background
field Eq.(\ref{nonabel}).
Even though it is not totally trustworthy, the loop
expansion can still provide indicative information.
Furthermore, to examine whether the qualitative
feature of the one-loop effective
potential strongly depends on the choice of the background
field is also interesting. In particular, we would like to
find out whether the coordinate singularity has any connection
with the existence of the imaginary part in the effective
potential. In this paper we will only address the question
of the one-loop
calculation and leave the lattice study for the future.

  In doing the one-loop calculation we encountered several
technical points, which were known in the literature but not
emphasized enough. We decide to briefly go through (in the
next section) the standard background field method in gauge
theories, with those technical points in mind. In section
III we explicitly calculate the one-loop effective potential
for the non-Abelian gauge background in the three dimensional
SU(2) Yang-Mills gauge theory. Unfortunately, we were unable
to carry out a similar calculation in four dimensions (except
the divergent part, which is done in the appendix), due to
technical difficulties in evaluating certain integrals.
Finally, we summarize and briefly discuss the generalization
of the lattice background field method in section IV.

\section{FORMALISM}

  In this section we outline the basic steps in the background
field method in gauge theories \cite{abbott}. There are two
reasons to go through the well-known method. The first is to
establish our own notations. The second is to emphasize two
technical aspects, the question of whether it is necessary to
require the background field satisfy the classical equation of
motion and the gauge choice in the evaluation of the functional
integral. When possible we follow the convention of Abbott
\cite{abbott}.

  The generating functional in the background field method
(in Euclidean space) starts with
\begin{equation}
{\tilde Z}[J,A]=\int[dQ]\,
\det\biggl[{\delta G^a\over\delta\omega^b}\biggr]
\exp\biggl\{-\int d^dx\bigl[{\cal L}(A+Q)
+J_\mu^a Q_\mu^a\bigr]\biggr\}
\prod_{x,a} \delta[G^a],
\label{zstart}
\end{equation}
where $G^a\equiv\partial_\mu Q_\mu^a+g f^{abc} A_\mu^b Q_\mu^c
=D_\mu^{ab}(A)Q_\mu^b$ is the background field gauge condition
and $\det\left[\delta G^a/\delta\omega^b\right]$ is the
corresponding Jacobian. It is clear from the above definition
the background field $A_\mu^a$ is fixed and should be regarded
as an external parameter in the process of doing the $Q$--integral.
Note also that we do not exponentiate the gauge constraint
$\prod_{x,a} \delta[G^a]$, in contrast with what was usually done.
The necessity of enforcing the gauge constraint with explicit
delta-functions in Eq.(\ref{zstart}) will be discussed
in detail soon.

  It is easy to verify that ${\tilde Z}[J,A]$ is invariant under
the infinitesimal gauge transformations
\begin{mathletters}
\label{ginv}
\begin{equation}
\delta A_\mu^a=-f^{abc}\omega^b A_\mu^c
+{1\over g}\partial_\mu \omega^a\, ;
\end{equation}
\begin{equation}
\delta J_\mu^a=-f^{abc}\omega^b J_\mu^c\, .
\end{equation}
\end{mathletters}
As a consequence of this invariance, the effective action
in the background field method,
defined as the Legendre transform of $\tilde Z$,
\begin{equation}
{\tilde \Gamma}[{\tilde Q},A]=-\ln {\tilde Z}[J,A]
+\int d^dx\,J_\mu^a {\tilde Q}_\mu^a,
\label{lt}
\end{equation}
with ${\tilde Q}_\mu^a=\delta \ln {\tilde Z}/ \delta J_\mu^a$,
is invariant under
\begin{mathletters}
\label{gt}
\begin{equation}
\delta A_\mu^a=-f^{abc}\omega^b A_\mu^c
+{1\over g}\partial_\mu \omega^a\, ;
\end{equation}
\begin{equation}
\delta {\tilde Q}_\mu^a=-f^{abc}\omega^b {\tilde Q}_\mu^c\, .
\end{equation}
\end{mathletters}
As shown by Abbott \cite{abbott}, when ${\tilde Q}_\mu^a=0$,
${\tilde \Gamma}[0,A]$ coincides with the usual effective action.
Since Eq.(\ref{gt}) is a pure gauge transformation when $\tilde Q$
vanishes, ${\tilde \Gamma}[0,A]$ must be a gauge invariant
functional of $A_\mu^a$.

  A standard way to evaluate ${\tilde Z}[J,A]$ explicitly is
to make a loop expansion. For the purpose of convenience, we
exponentiate the Jacobian $\det[\delta G^a/ \delta\omega^b]$
and gauge constraints $\prod_{x,a} \delta[G^a]$ by introducing
the Faddeev-Popov ghost fields ($\theta^a$ and $\bar\theta^a$)
and a real scalar auxiliary field $\sigma^a$ respectively,
\begin{eqnarray}
\tilde{Z}[J,A] &=&\int[dQ][d\theta][d\bar\theta][d\sigma]\,
\exp\biggl\{-\int d^dx\bigl[{\cal L}(A+Q)
+{\bar\theta^a}
D_\mu^{ac}(A)D_\mu^{cb}(A)\theta^b + \nonumber \\
& &+2i\sigma^a D_\mu^{ab}(A)Q_\mu^b
+J_\mu^a Q_\mu^a\bigr]\biggr\}\, .
\label{zend}
\end{eqnarray}
If we only want to calculate $\tilde Z[J,A]$ to one-loop order,
then it is equivalent to evaluate Eq.(\ref{zend})
in the steepest descent approximation,
\begin{eqnarray}
\tilde{Z}[J,A]&\approx& {\tilde Z}_1[J,A]\equiv
\int[dQ][d\theta] [d\bar\theta][d\sigma]
\exp\biggl\{-\int d^dx\bigl[{\cal L}(A)
+{\cal L}^{(1)}(A,Q) + \nonumber \\
& & {\cal L}^{(2)}(A,Q)
+{\bar\theta^a} D_\mu^{ac}(A)D_\mu^{cb}(A)\theta^b
+2i\sigma^a D_\mu^{ab}(A)Q_\mu^b
+J_\mu^a Q_\mu^a\bigr]\biggr\}\, ,
\label{z1}
\end{eqnarray}
where
\begin{equation}
{\cal L}^{(1)}(A,Q)=-Q_\mu^a D_\mu^{ab}(A)F_{\mu\nu}^b(A)\, ;
\end{equation}
\begin{equation}
{\cal L}^{(2)}(A,Q)=Q_\mu^a\biggl[
-{1\over 2}\left(D_\rho(A)D_\rho(A)\right)^{ab}\delta_{\mu\nu}
+ig F_{\mu\nu}^{ab}(A)\biggr]Q_\nu^b
\equiv Q_\mu^a M_{\mu\nu}^{ab}(A) Q_\nu^b\, .
\label{ma}
\end{equation}
In arriving at the expression of ${\cal L}^{(2)}(A,Q)$, the gauge
constraint condition $D^{ab}_\mu Q_\mu^b=0$ has been used.

  We can now do the $Q$--integral in ${\tilde Z}_1[A,J]$ and
make the Legendre transform as in Eq.(\ref{lt}). By setting
${\tilde Q}_\mu^a=0$ the final expression for the effective
action to one-loop order can be written as
$\Gamma[A]=-\ln {\tilde Z}_1[A]$, with
\begin{eqnarray}
{\tilde Z}_1[A]&=&e^{-\int d^dx {\cal L}(A)}\cdot
\det\bigl[-D(A)D(A)\bigr] \cdot \nonumber \\
& & \int [dQ][d\sigma]
\exp\biggl\{-\int d^dx\bigl[Q_\mu^a M_{\mu\nu}^{ab}
(A)Q_\nu^b + 2i\sigma^a
D_\mu^{ab}(A)Q_\mu^b\bigr]\biggr\}  \nonumber \\
&=&e^{-\int d^dx {\cal L}(A)}\cdot
\det\bigl[-D(A)D(A)\bigr] \cdot
\biggl\{\det M_{\mu\nu}^{ab}\cdot\det N^{ab}\biggr\}^{-1/2} \, ,
\label{aeff}
\end{eqnarray}
with $N^{ab}=-D_\mu^{ac} (M^{-1})^{ce}_{\mu\nu}D_\nu^{eb}$.
The first factor is the classical contribution. The second and
third factors are the one-loop quantum corrections due to the
ghost field ($\theta$) and fluctuation field ($Q$), respectively.
Notice that the linear term ${\cal L}^{(1)}(A,Q)$ drops out
automatically in the calculation process. There is no need to
require the background field to satisfy the classical equation
of motion $D_\mu^{ab}(A)F_{\mu\nu}^b(A)=0$. This independence of
the linear term remains true to all orders in loop expansion,
because the effective potential is a sum of all one-particle
irreducible diagrams with $A$ fields on the external lines and
$Q$ fields on internal lines (when ${\tilde Q}_\mu^a=0$). In
fact, the linear term is always compensated by the source $J_\mu^a$.
Remember that the physical limit in the background field method
is ${\tilde Q}_\mu^a=0$, not $J_\mu^a=0$.

  At this point we would like to comment on the implementation of
the gauge condition $\prod_{x,a}\delta[G^a]$ in Eq.(\ref{zstart}).
The standard approach is to exponentiate this factor,
\begin{equation}
{\tilde Z}[J,A;\alpha]=\int[dQ]\,
\det\biggl[{\delta G^a\over\delta\omega^b}\biggr]
\exp\biggl\{-\int d^dx\bigl[{\cal L}(A+Q)
-{1\over 2\alpha}(G^a)^2+J_\mu^a Q_\mu^a\bigr]\biggr\},
\label{zexp}
\end{equation}
where $\alpha$ is the gauge fixing parameter. In order to write
Eq.(\ref{zstart}) into the form of Eq.(\ref{zexp}), one has to
generalize the gauge condition as
$\prod_{x,a}\delta[G^a-f^a]$, with $f^a$ being arbitrary,
and then to integrate out $f^a$ with the weighting factor
$\exp[-(f^a)^2/2\alpha]$. If ${\tilde Z}[J,A]$ with the
generalized gauge condition were independent of $f^a$, as in
the case of zero background field $A_\mu^a=0$,
${\tilde Z}[J,A;\alpha]$ would be independent of the gauge
parameter $\alpha$ and therefore the exact equivalence between
Eq.(\ref{zstart}) and Eq.(\ref{zexp}). However, in the presence
of non-vanishing background field $A_\mu^a$, ${\tilde Z}[J,A]$
in general would depend on $f^a$, as can be easily seen in
Eq.(\ref{zend}) by adding a term $2i\sigma^a f^a$ to the
exponential. Physically, this dependence on $f^a$ is natural,
because there would exist correlations between $f^a$
and certain combinations of the background field, such as
$(D_\mu^{ab}(A)f^b)^2$ when $A_\mu^a$ is non-vanishing.
If indeed ${\tilde Z}[J,A]$ depended on $f^a$, then the integral
\begin{equation}
{\tilde Z}[J,A;\alpha]\equiv
\int [df] \,{\tilde Z}[J,A] \bigg|_{G^a=f^a}\,
\cdot\exp\biggl\{-{1\over 2\alpha}(f^a)^2\biggr\}
\end{equation}
would have to depend on $\alpha$ in general, which in turn implies
that Eq.(\ref{zexp}) could not be equivalent to Eq.(\ref{zstart})
at all times.

  This gauge dependence of the effective action calculation
has long been recognized in the literature \cite{vilkovisky}.
It was argued by Vilkovisky that the correct choice of the
gauge parameter is to take $\alpha\rightarrow 0$ limit, or to
choose the Landau background field gauge.
Since $\exp[-x^2/2\alpha]\propto\delta(x)$ in the limit of
$\alpha=0$, Eq.(\ref{zexp}) is formally equivalent to
Eq.(\ref{zstart}) in the Landau background field gauge. It
should be emphasized that Eq.(\ref{zexp}) would still define
a gauge invariant effective action with respect to the
background field $A_\mu^a$. It is only the functional
form of the effective action that depends on the gauge
parameter when $\alpha\ne 0$. Finally, in the context of finite
temperature QCD, this gauge dependence has been emphasized by
Hansson and Zahed \cite{hansson}.

\section{EFFECTIVE POTENTIAL}

  In this section we apply the background field method to
calculate the effective potential for the background field
mentioned in the Introduction in three dimensional SU(2)
Yang-Mills gauge theory in Euclidean space. The three dimensional
analog of Eq.(\ref{nonabel}) would be,
\begin{equation}
A_\mu^a(x)=h\delta_{a\mu}\, .
\label{ba}
\end{equation}
This choice of $A_\mu^a$ leads to non-vanishing field strengths,
through commutator terms rather than derivative terms as in the
case of Savvidy's ansatz,
\begin{equation}
B_\mu^a\equiv {1\over 2}\epsilon_{\mu\alpha\beta}
F_{\alpha\beta}^a(x)=gh^2\delta_{a\mu}\, ,
\label{bb}
\end{equation}
with $g$ being the coupling constant. Due to the non-trivial
correlation between the color and space-time orientations
the determinant $\det B_\mu^a\ne 0$ when $h\ne 0$.
Notice that in Euclidean space all components of the field
strength are magnetic-like.

  Since the background field is a constant in space-time it
is convenient to work in momentum-space. Using the formula
$\det M=\exp({\rm Tr}\ln M)$ combined with Eq.(\ref{aeff}),
we have the expression for the effective potential to
one-loop order in our ansatz
\begin{equation}
V_{\rm eff}(h)={\cal L}(h)
-\int{d^3p\over(2\pi)^3}\ln G(h;p)+{1\over2}\int{d^3p\over(2\pi)^3}
\biggl\{\ln M(h;p)+\ln N(h;p) \biggr\} \, ,
\label{veff}
\end{equation}
where
\begin{equation}
G(h;p)\equiv \det\bigl[-D_\mu^{ac}(h;p)D_\mu^{cb}(h;p)\bigr]
=(p^2+2m^2)(p^4+4m^4)\, ;
\end{equation}
\begin{equation}
M(h;p)\equiv \det\bigl[M_{\mu\nu}^{ab}(h;p)\bigr]
=p^6(p^2-4m^2)(p^4+6m^2p^2+16m^4)(p^6+4m^2p^4-16m^6)\, ;
\label{appa}
\end{equation}
\begin{eqnarray}
N(h;p) &\equiv & \det\bigl
[-D_\mu^{ac}(h;p)(M^{-1})_{\mu\nu}^{ce}(h;p)
D_\nu^{eb}(h;p)\bigr] \nonumber \\
&=&{(p^6-4m^6)(p^4+2m^2p^2+4m^4)
\over(p^4+6m^2p^2+16m^4)(p^6+4m^2p^4-16m^6)}\, ;
\end{eqnarray}
with $m\equiv gh$.
Non-positive definiteness of $M(h;p)$ and $N(h;p)$ for small
$p^2$ indicates that the background field in Eq.(\ref{ba})
is only a saddle point in configuration space and therefore
is not stable. To reveal the physical origin of the existence
of the unstable modes, let us explicitly write
\begin{equation}
M_{\mu\nu}^{ab}(h;p)={1\over2}
\bigg[ (p^2+2g^2h^2)\delta^{ab}\delta_{\mu\nu}
+2igh\epsilon^{abc}p_c\delta_{\mu\nu}
+2g^2h^2\epsilon^{abc}\epsilon_{c\mu\nu}\biggr]\, ,
\label{mh}
\end{equation}
which is a $9\times9$ matrix for a given momentum $p$. This
matrix can be interpreted as the Hamiltonian of a relativistic
spin-1 ($\vec{\sigma}_s$) and color spin-1 ($\vec{\sigma}_c$)
boson, with the first term being the free particle part, the
second the $\vec{p}\cdot\vec{\sigma}_c$ and the third
$\vec{\sigma}_s\cdot\vec{\sigma}_c$. The last two terms can be
negative for low momentum, depending on the relative
orientations of $\vec{p}$,
$\vec{\sigma}_s$ and $\vec{\sigma}_c$. It is interesting to
observe that in the case of Savvidy's ansatz the corresponding
$M$ matrix is the Landau diamagnetic Hamiltonian.

  Using the regularization of Salam and Strathdee \cite{salam},
which is a variation of Schwinger's proper time method
\cite{schwinger}, the integrals in Eq.(\ref{veff}) can be
worked out explicitly. Because the three dimensional
Yang-Mills theory is super renormalizable, the only divergence
we encounter is an overall additive constant. The regularization
procedure includes three steps. Firstly, an integral
representation for logarithmic function is used. For real $E$,
positive or negative,
\begin{equation}
\ln(E-i\delta)={1\over\epsilon}
-{i^\epsilon\over\epsilon\Gamma(\epsilon)}
\int dt\, t^{\epsilon-1}\, e^{-it(E-i\delta)}\, ,
\label{ssr}
\end{equation}
in the limit of $\epsilon\rightarrow 0^+$. The $i\delta$
in Eq.(\ref{ssr}) with $\delta\rightarrow 0^+$ is to ensure
the convergence of the integral. When $E$ is explicitly
complex, Eq.(\ref{ssr}) can be easily generalized. Then the
momentum integration can be done using
\begin{equation}
\int {d^3p\over(2\pi)^3}e^{-itp^2}={1\over 8\pi^{3/2}}
(it)^{-3/2}\, .
\end{equation}
The remaining $t$--integral can be converted into a Gamma
function through the contour integral technique. Finally,
the limit of $\epsilon\rightarrow 0$ is taken.

  After dropping an overall divergent constant, the
effective potential has the following expression,
\begin{mathletters}
\label{final}
\begin{equation}
V_{\rm eff}(h)={3\over2}g^2h^4-{2-\sqrt{2}\over3\pi}g^3h^3
-i{5\over6\pi}g^3h^3\, ,
\end{equation}
or if we define $B\equiv gh^2$,
\begin{equation}
V_{\rm eff}(B)={3\over2}B^2-{2-\sqrt{2}\over3\pi}(gB)^{3/2}
-i{5\over6\pi}(gB)^{3/2}\, .
\end{equation}
\end{mathletters}
Again the first term in Eq.(\ref{final}) is the classical
contribution, the second term is the ghost field ($\theta$)
contribution and the third is the fluctuation field ($Q$)
contribution. Compared to the calculation of Trottier
\cite{trottier} in the Savvidy ansatz, we find qualitative
similarity between the two results. If we ignore the imaginary
part in the effective potential temporarily, we would find a
non-trivial minimum,
\begin{mathletters}
\label{min}
\begin{equation}
h_{\rm min}={2-\sqrt{2}\over 6\pi} g\, ,
\end{equation}
or
\begin{equation}
B_{\rm min}=\biggl({2-\sqrt{2}\over 6\pi}\biggr)^2 g^3\, .
\end{equation}
\end{mathletters}
Since the effective potential evaluated at the minimum point
is the energy density relative to the perturbative vacuum,
$V_{\rm eff}(h_{\rm min})<0$ signals that the system prefers
to spontaneously generate color magnetic fields in order
to gain energy and in turn signals the instability
of the perturbative vacuum. However, due to the presence of
the imaginary part in $V_{\rm eff}$, the vacuum characterized
by Eq.(\ref{min}) itself can not be stable either, at least
not in the loop expansion to one-loop order.

  In passing, if the Feynman gauge ($\alpha=1$) were used,
which would be equivalent to ignoring the term $\ln N(h;p)$
in Eq.(\ref{veff}), we would have obtained
\begin{equation}
V_{\rm eff}(h)\bigg|_{\alpha=1}={3\over2}g^2h^4-C_R g^3h^3
-i C_I g^3h^3\, ,
\end{equation}
with $C_R=0.43514...$ and $C_I=0.26989...$.
It is possible to get exact expressions for $C_R$ and $C_I$.
But these expressions are complicated and not illuminating. We
will not bother to write them down here. Nevertheless, it is
sufficient to mention that the
values of $C_R$ and $C_I$ are different from their corresponding
counterparts in Eq.(\ref{final}).

  To generalize the above calculation to four dimensions, we
could try the ansatz $A_i^a=h\delta_{ia}$ ($i=1,2,3$) in
the temporal gauge $A_0^a=0$, as in Eq.(\ref{nonabel}).
In fact, such a calculation
already exists in the literature \cite{singer} about
one decade ago, with a different motivation.
However, the minimum point found in that
work is not renormalization group invariant, which in turn
implies that their $\beta$--function differs from the
correct SU(2) value (by a factor of 2). This is in contradiction
with rather general arguments, for example in \cite{maiani},
that the ultraviolet behaviors of the Yang-Mills theory stays
the same, independent of the presence or absence of
a background field, since the background field is only relevant
in the infrared regime. Furthermore, in the work of L\"{u}scher,
in terms of Hamiltonian approach \cite{luscher}, and of Kolker
and van Baal, in terms of Lagrangian approach \cite{vanbaal},
where they calculated the coefficients of the effective potential
in SU(2) Yang-Mills theory for constant background field up to
fourth order in the background field on a finite torus, the
$\beta$--function was found not to be modified by the presence
of the background field. The same conclusion was shown
\cite{mluscher} to hold explicitly at one loop level, at least
for some classes of background fields, without expanding the
effective potential as a power series of the background fields.
 Since the $\beta$--function is gauge
(or $\alpha$) independent in the dimensional regularization
and minimal subtraction scheme \cite{gross}, although the finite
part is explicitly $\alpha$ dependent, we suspect that the
authors of reference \cite{singer} overlooked certain subtleties
in calculating the four dimensional functional determinant.
Due to the lack of details in their paper we could not explicitly
check their results. However, since the imaginary part of
the effective potential is finite and easy to calculate, we have
verified that the authors of \cite{singer} missed a term due to
one of their $\lambda_{4,5,6,7}$ in Eq.(5), which could become
negative for low enough momenta. To find out what really happens
in our case, we will show, in the appendix,
that the $\beta$--function is not affected in the
presence of the background field Eq.(\ref{nonabel}) through
an explicitly calculation of the divergent part of the effective
potential.

\section{SUMMARY AND DISCUSSION}

  To summarize, we have calculated the effective potential for
a particularly chosen non-Abelian background field in the three
dimensional SU(2) Yang-Mills theory. Technical questions related
to the linear term and the gauge choice were illustrated.
The result is found to be
qualitatively similar to that of the Abelian ansatz of the Savvidy
type \cite{trottier}, both in real part, which indicates a
spontaneous generation of the color magnetic field, and imaginary
part, which signals the instability of the background field
Eq.(\ref{ba}) as the vacuum configuration under the loop expansion.
Though not explicitly calculated, the four dimensional effective
potential with the same non-Abelian ansatz in the temporal gauge
is expected to behave very much the same. Given the qualitative
similarity between the three dimensional and four dimensional
effective potential in the Savvidy ansatz, we suspect that the
effective potential is insensitive to the coordinate singularity,
${\rm Det}B_i^a=0$, if it indeed exists.

  It is important to recognize that the starting point in
Eq.(\ref{zstart}) would be a well defined problem if the functional
were evaluated non-perturbatively. The appearance of the imaginary
part in the effective potential is only caused by the loop expansion.
In other words, when we were doing steepest descent approximation,
we were expanding at a saddle point. In addition, even if the
expansion point were a true minimum, the stationary solution of the
effective potential calculated up to a finite loop order could not
be trusted quantitatively, due to the fact that at the stationary
point of the effective potential the higher order terms become as
important as lower order terms and hence the loop expansion breaks
down. Therefore the result in this paper could at best be regarded
as indicative. The final answer has to be settled by a non-perturbative
means, such as the lattice simulation as mentioned in the Introduction.

  In fact, a lattice generalization of the background field method
is rather straightforward. Let us consider
\begin{equation}
Z_L[J_\mu(x),B_\mu(x)]\equiv\int dU_\mu(x)\,
\exp\biggl\{-S[U_\mu(x)B_\mu(x)]+{\rm Tr}J_\mu(x) f[U_\mu(x)]\biggr\}\, ,
\end{equation}
where $U_\mu(x)$ is the standard link variable, $B_\mu(x)$ is the
background link variable, $S$ is the usual Wilson lattice action,
$J_\mu(x)$ is a matrix valued external current and $f[U]$ is an
arbitrary function satisfying $gf[U]g^\dagger=f[gUg^\dagger]$ for
any unitary matrix $g$. Using the property of the invariance
under an unitary transformation for the Haar measure one can easily
verify the following
\begin{equation}
Z_L[\tilde{J}_\mu(x),\tilde{U}_\mu(x)]
\bigg|_{\tilde{U}_\mu(x)=g(x)U_\mu(x)g^\dagger(x+\mu),\,
        \tilde{J}_\mu(x)=g(x)J_\mu(x)g^\dagger(x)}
=Z_L[J_\mu(x),U_\mu(x)]\, ,
\end{equation}
a lattice version of Eq.(\ref{ginv}). A Legendre transform of
$Z_L[J_\mu,U_\mu]$ would lead to a gauge invariant effective action,
provided that the induced gauge field is constrained to have zero
expectation value by adjusting $J_\mu(x)$, just as in the continuum
case. The remarkable thing here is that we do not need to fix the
gauge on a lattice and therefore the resulted effective action is
unique for a given choice of the functional form of $f$.

  As mentioned in the Introduction the background field ansatz
Eq.(\ref{ba}) can be conveniently realized on a lattice. The constant
nature of the ansatz avoids problems with the boundary condition
and non-uniformness of the lattice constant effect due to the
linear rising ansatz of Savvidy. Since Eq.(\ref{ginv}) only involves
one parameter it may not be difficult to find a way to adjust the
external current $J_\mu$ to ensure a vanishing of the expectation
value for the induced quantum field. Work along this line will
be pursued in the future.

\acknowledgements

It is our pleasure to thank H.~B.~Nielsen, H.~Trottier, P.~van Baal,
and especially Ken~Johnson and Janos~Polonyi for many very fruitful
suggestions and discussions. This work is supported in part by funds
provided by the U. S. Department of Energy (D.O.E.) under contracts
\#DE-FG06-88ER40427, \#DE-AC02-76ER03069 and \#DE-FG02-91ER40676.

\appendix
\section*{}

In this appendix we explicitly calculate the
divergent part of the effective potential for the background
field Eq.(\ref{nonabel}) and
show that the $\beta$--function remains the same as calculated
with zero background field in four dimensions. Since we do not
expect the divergent part to depend on the gauge choice in the
Minimal subtraction scheme \cite{gross}, we will pick the
Feynman gauge ($\alpha=1$) for convenience.

The four-dimensional analog of $M(h;p)$, Eq.(\ref{appa}), in the
background field Eq.(\ref{nonabel}), is given by
\begin{equation}
M(h;p)=\det[M_{00}^{ab}(h;p)]\cdot\det[M_{ij}^{ab}(h;p)]\, ,
\end{equation}
with (again $m\equiv gh$)
\begin{eqnarray}
\det[M_{00}^{ab}(h;p)]&=&[p^2+2m^2]\,
[(p^2+2m^2)^2-4m^2{\bf p}^2)]\, , \\
\det[M_{ij}^{ab}(h;p)]&=&[p^4-4m^2{\bf p}^2]\,
[(p^4+4m^2p^2)^2-4m^2{\bf p}^2(p^2+2m^2)^2]
\times \nonumber \\
& &\times [p^6+10m^2p^4-4m^2{\bf p}^2p^2
+24m^4p^2-8m^4{\bf p}^2]\, ,
\end{eqnarray}
while the ghost determinant
$G(h;p)=\det[M_{00}^{ab}(h;p)]$.
To regularize the momentum integration we use the
finite-temperature extension of
the usual dimensional regularization
\begin{equation}
\int[dp]\, f(p) \equiv {1\over\beta}\mu^{2\epsilon}
\int {d^{3-2\epsilon}{\bf p}\over(2\pi)^{3-2\epsilon}}
\left[f(p_0=0,{\bf p}^2)+2\sum_{n=1}^\infty
f(p_0=\omega_n\equiv{2\pi n\over\beta},{\bf p})\right]\, .
\end{equation}

Since an overall additive constant of $V_{\rm eff}(h)$ is
irrelevant we can write
\begin{equation}
V_{\rm eff}(h)\equiv \int_0^{m^2} dm^2\,
{\partial V_{\rm eff}(h)\over\partial m^2}\, .
\end{equation}
To carry out the spatial momentum integration
it is convenient to rewrite
\begin{eqnarray}
\det[M_{00}^{ab}(h;p)]&=&\prod_{l=1}^3 \,
[{\bf p}^2+a_l(m^2,\omega_n^2)]\, , \\
\det[M_{ij}^{ab}(h;p)]&=&\prod_{l=4}^{12} \,
[{\bf p}^2+a_l(m^2,\omega_n^2)]\, ,
\end{eqnarray}
where the $a_l$'s can be solved from Eq.(5.2) and (5.3). The
explicit expressions of $a_l$'s are too lengthy to be presented
here. However, if our interest is to calculate the divergent
part (or $1/\epsilon$ term) we only need to notice that $a_l$'s
can be expanded as a power series of $m$ (when $n\ne 0$).
In terms of $a_l$'s the one-loop part of
the effective potential becomes
\begin{equation}
{\partial V^{\rm 1-loop}_{\rm eff}(h)\over\partial m^2}=
{1\over 2}\sum_{l=1}^{12} s_l
\int[dp] \, {1\over {\bf p}^2+a_l(m^2,\omega_n^2)} \,
{\partial a_l(m^2,\omega_n^2)\over \partial m^2}\, ,
\end{equation}
with $s_l=-1$ for $l=1,2,3$ and $s_l=+1$ for $l=4,5,...,12$.
The spatial momentum integration can be done trivially,
yielding
\begin{equation}
{\partial V^{\rm 1-loop}_{\rm eff}(h)\over\partial m^2}=
{\mu^{2\epsilon}\Gamma(\epsilon-1/2)\over
2\beta(4\pi)^{3/2-\epsilon}}\sum_{l=1}^{12}
{s_l\over 3/2-\epsilon}{\partial\over\partial m^2}
\left[a_l(m^2,0)^{3/2-\epsilon}+2\sum_{n=1}^\infty
a_l(m^2,\omega_n^2)^{3/2-\epsilon}\right]\, .
\end{equation}

To isolate the pole term in $\epsilon$ we need to expand
$a_l(m^2,\omega_n^2)^{3/2-\epsilon}$ in a power series of
$m/\omega_n$ in the infinite Matsubara frequency sum,
\begin{equation}
a_l(m^2,\omega_n^2)^{3/2-\epsilon}=\omega_n^{3-2\epsilon}
\left[1+\sum_{i=1}^\infty a_l^{(i)}
\left({m\over\omega_n}\right)^i\right]\, ,
\end{equation}
because $a_l(0,\omega_n)=\omega_n^2$. In general,
the $a_l^{(i)}$'s are complex, but analytic in $\epsilon$.
By noticing the fact that the Riemann zeta function
$\zeta(s)\equiv \sum_{n=1}^\infty n^{-s}$ is singular
only at $s=1$, we merely need to know $a_l^{(4)}$,
which corresponds to $\omega_n^{-1-2\epsilon}$ terms. An
explicit calculation gives $a_1^{(4)}=3/2$,
$a_2^{(4)}=a_3^{(4)}=-9/8$, $a_4^{(4)}=a_5^{(4)}=15/8$,
$a_6^{(4)}=a_7^{(4)}=3/2$, $a_8^{(4)}=a_9^{(4)}=15/8$,
$a_{10}^{(4)}=-9/2$ and $a_{11}^{(4)}=a_{12}^{(4)}=39/8$.
So the expression for the divergent part of the effective
potential becomes
\begin{eqnarray}
V_{\rm eff}^{\rm 1-loop, div}(h) &=&
{\mu^{2\epsilon}\Gamma(\epsilon-1/2)\over
\beta(4\pi)^{3/2-\epsilon}}\sum_{l=1}^{12}
{s_l a_l^{(4)}\over 3/2-\epsilon} \, m^4
\left({\beta\over 2\pi}\right)^{1+2\epsilon}
\zeta(1+2\epsilon) \nonumber \\
&=& -{g^4h^4\over 24\pi^2\epsilon}
\sum_{l=1}^{12}s_l a_l^{(4)}\, ,
\end{eqnarray}
which is independent of $\beta$ (temperature) and real, as
it should be. Using the explicit values of $a_l^{(4)}$'s and
adding the classical contribution $3g^2h^4/2$, we finally have
\begin{equation}
V_{\rm eff}(h)={3\over 2}g^2 h^4
\left[1-{11g^2\over 24\pi^2\epsilon}\right]
+\, {\rm finite}\,\,{\rm terms}\, .
\end{equation}
Due to the explicit gauge invariance of the effective potential
in the background field approach, the product $gh$ needs no
renormalization \cite{abbott}. Hence the factor in the bracket in
the above equation is nothing but $Z_g^2$, which leads to the
correct $\beta$--function. Therefore, the ultraviolet behavior
of the SU(2) Yang-Mills theory in the background
Eq.(\ref{nonabel}) remains universal.

\end{document}